\documentclass[11pt]{article}
\usepackage{graphicx}
\usepackage{float} 
\usepackage{amsmath}
\usepackage{amsfonts}   
\usepackage{amssymb}

\def\ph{\phi}

\begin{document}
\date{}
\title{{\bf{\Large Effective Values of Komar Conserved Quantities and Their Applications}}}
\author{
 {\bf {\normalsize Sujoy Kumar Modak}$
$\thanks{e-mail: sujoy@bose.res.in}}\\
 {\normalsize S.~N.~Bose National Centre for Basic Sciences,}
\\{\normalsize JD Block, Sector III, Salt Lake, Kolkata-700098, India}
\\[0.3cm]
{\bf {\normalsize Saurav Samanta}$
$\thanks{e-mail: srvsmnt@gmail.com}}\\
 {\normalsize Narasinha Dutt College}
\\{\normalsize 129, Belilious Road, Howrah-711101, India}
\\[0.3cm]
}

\maketitle


\begin{abstract}
 
We calculate the effective Komar angular momentum for the Kerr-Newman (KN) black hole. This result is valid at any radial distance on and outside the black hole event horizon. The effcetive values of mass and angular momentum are then used to derive an identity ($K_{\chi^{\mu}}=2ST$) which relates the Komar conserved charge ($K_{\chi^{\mu}}$) corresponding to the null Killing vector ($\chi^{\mu}$) with the thermodynamic quantities of this black hole. As an application of this identity the generalised Smarr formula for this black hole is derived. This establishes the fact that the above identity is a local form of the inherently non-local generalised Smarr formula. 
    
\end{abstract}

\section{Introduction}
For a spacetime endowed with symmetries, one can define Killing vectors corresponding to each of these symmetry directions. One major application of these Killing vectors is to find the constants associated with the motion along some geodesic. This is often done by exploiting the Komar expressions of conserved quantities\cite{komar} which can be written in a covariant form. For example, the Kerr-Newman spacetime has two Killing vectors $\xi^{\mu}_{(t)}$ and $\xi^{\mu}_{(\phi)}$ ($t$ being the time axis and $\phi$ is the azimuthal angle). Therefore one has two conserved quantities, namely mass ($M$) and angular momentum ($J$) corresponding to these Killing vectors. By construction, mass and angular momentum of a asymptotically flat spacetime are defined with respect to an observer situated far away from the horizon and not influenced by the spacetime curvature. However, as one approaches towards the event horizon the surrounding spacetime no longer remains flat and therefore the earlier results get modified. 

Indeed the effective mass (conserved quantity corresponding to $\xi^{\mu}_{(t)}$ which is asymptotically timelike) of the Kerr-Newman black hole, as calculated by Cohen and DeFelice \cite{cohen2}, is given by
\begin{eqnarray}
M_{\text{eff}}=M-\frac{Q^2}{2r}-\frac{Q^2(r^2+a^2)}{2ar^2}\tan^{-1}\left(\frac{a}{r}\right).
\label{meff}
\end{eqnarray}    
This result is also consistent with the observed value of mass at asymptotic infinity, as,  $\displaystyle\lim_{r\to\infty}M_{\text{eff}}=M$. For $a=0$ we get  the Reissner-Nordstrom black hole for which $M_{\textrm{eff}}^{RN}=M-\frac{Q^2}{r}$ \cite{Carrol} and for the chargeless case (Kerr black hole), $M_{\text{eff}}=M$ is satisfied at any distance from the event horizon. Both the Reissner-Nordstrom and the Kerr-Newman black hole, when reduced to the Schwarzschild case, give the same limiting value of the effective mass  $M_{\textrm{eff}}^{Sch}=M$. Comparing with the Schwarzschild case, it is evident from (\ref{meff}) that there are two extra terms in the expression of the effective mass of the Kerr-Newman black hole which come from the charge ($Q$) and the reduced angular momentum ($a=\frac{J}{M}$). One term is purely elctrostatic energy ($\frac{Q^2}{r}$) contained within a sphere of radiuas $r$. There is another term where the reduced angular momentum is coupled with the electric charge of the black hole. This term vanishes both for chargeless and irrotational black holes. This coupling term is often called a gravito-electric effect. Both these contributions vanish at infinity and one recovers the asymptotic value of mass.  

In this paper we first evaluate the Komar integral corresponding to the vector $\xi^{\mu}_{(\phi)}$ at a finite radial distance for Kerr-Newman spacetime. This integral is performed at the boundary of a spatial 3-hypersurface. This boundary is chosen to be consists of time-synchronised events only. Due to this choice we are able to calculate the Komar integral at a finite $r$ for this non-static spacetime. The result for effective angular momentum follows from this calculation. Then using the effective values of Komar conserved charges we derive a new identity $K_{\chi^{\mu}}=2ST$ at the event horizon. Here the left hand side is the Komar conserved quantity corresponding to a null Killing vector $\chi^{\mu}$ and $S,T$ are black hole entropy and Hawking temperature respectively. We also connect this identity with a similar relationship $E=2ST$ ($E$ is the Noether charge of the diffeomorphism symmetry) which is found to be satisfied at any timelike Killing horizon of a static spacetime \cite{paddy1,paddy2,Padmanabhan:2009vy,Banerjee:2010yd,Banerjee:2010rx}. This horizon infact need not be a black hole event horizon, it can also be of Rindler type. However the validity of such an identity for a nonstatic spacetime is not obvious. Nevertheless our work suggests the validity of this identity with appropriate interpretation of $E$ in terms of Komar conserved quantities. As a byproduct of this identity we derive the generalised Smarr formula for the Kerr-Newman black hole. Finally we explain the use of effective expressions of Komar integrals in finding the coefficient of the logarithmic correction to semiclassical black hole entropy for the Kerr-Newman black holes.

\section{Effective values of Komar conserved charges for the Kerr-Newman metric}
The spacetime metric for the Kerr-Newman black hole, in Boyer and Lindquist coordinates, is given by
\begin{eqnarray}
ds^2=g_{ij}dx^idx^j=-\frac{\Delta\Sigma}{A}dt^2+\frac{A{\textrm{sin}}^2\theta}{\Sigma}(d\phi-\Omega dt)^2+\frac{\Sigma}{\Delta}dr^2+\Sigma d\theta^2
\end{eqnarray}
where
\begin{eqnarray}
\Omega&=&-\frac{a}{A}(Q^2-2Mr)\label{om}\\
\Sigma&=&r^2+a^2{\textrm{cos}}^2\theta\\
\Delta&=&r^2+a^2+Q^2-2Mr\\
A&=&\Sigma(r^2+a^2)-a^2(Q^2-2Mr){\textrm{sin}}^2\theta.
\end{eqnarray}
Now to find the conserved quantities for this space time one needs to use the Komar expressions. There are two expressions for two Killing vectors--one leads to mass and other gives the angular momentum. The effective Komar mass, in a coordinate free notation is defined as \cite{cohen2,wald,dadhich1,dadhich2}
\begin{eqnarray}
K_{\xi^{\mu}_{(t)}}=M_{\textrm{eff}}=-\frac{1}{8\pi}\int_{\partial \Sigma}{}^*d\sigma,
\label{eeff}
\end{eqnarray}
where $K_{\xi^{\mu}_{(t)}}$ is the Komar conserved quantity corresponding to the time-Killing one form $\sigma =\xi_{(t)\mu}dx^{\mu}=g_{0\mu}dx^{\mu}=g_{00}~dt+g_{03}~d\ph$ and $^*d\sigma$ is the dual to the two form $d\sigma$. 
Using this, the integration (\ref{eeff}) was computed earlier in \cite{cohen2} to find the effective mass (\ref{meff}). 

Following a method similar as \cite{cohen2}, we now proceed to calculate the effective angular momentum for the Kerr-Newman spacetime, as measured by a local observer situated at some arbitrary radial distance $r$. 

The Komar definition of the conserved quantity, corresponding to the spacelike Killing vector $\xi^{\mu}_{(\phi)}$, in a coordinate free notation is given by
\begin{eqnarray}
K_{\eta}=-\frac{1}{8\pi}\int_{\partial \Sigma}{}^*d\eta
\label{peff}
\end{eqnarray}
where the spacelike  Killing one form is defined as
\begin{eqnarray}
\eta&=\xi_{(\phi)\mu}dx^{\mu}= g_{\mu 3}dx^{\mu} = g_{03}(r,\theta)dt+g_{33}(r,\theta)d\phi.
\label{111}
\end{eqnarray}
Differentiating the above equation we find 
\begin{eqnarray}
d\eta=\frac{\partial g_{03}}{\partial r}dr\wedge dt+\frac{\partial g_{03}}{\partial\theta}d\theta\wedge dt+\frac{\partial g_{33}}{\partial r}dr\wedge d\ph+\frac{\partial g_{33}}{\partial\theta}d\theta\wedge d\phi
\label{deta}
\end{eqnarray}
 Instead of working with $dt,dr,d\theta,d\phi$ it is important to work with the following orthonormal one forms \cite{cohen2}
\begin{eqnarray}
\hat{\varkappa}_0&=&-\left(\frac{\Delta\Sigma}{A}\right)^{\frac{1}{2}}dt\nonumber\\
\hat{\varkappa}_1&=&\left(\frac{\Sigma}{\Delta}\right)^{\frac{1}{2}}dr\nonumber\\
\hat{\varkappa}_2&=&\Sigma^{\frac{1}{2}}d\theta\nonumber\\
\hat{\varkappa}_3&=&\left(\frac{A{\textrm{sin}}^2\theta}{\Sigma}\right)^{\frac{1}{2}}\left(d\ph-\Omega dt\right)\nonumber
\label{ford}
\end{eqnarray}
 Using the inverse relations 
\begin{eqnarray}
dt&=&-\left(\frac{A}{\Delta\Sigma}\right)^{\frac{1}{2}}\hat{\varkappa}_0\nonumber\\
dr&=&\left(\frac{\Delta}{\Sigma}\right)^{\frac{1}{2}}\hat{\varkappa}_1\nonumber\\
d\theta&=&\left(\frac{1}{\Sigma}\right)^{\frac{1}{2}}\hat{\varkappa}_2\nonumber\\
d\phi&=&\left(\frac{\Sigma}{A{\textrm{sin}}^2\theta}\right)^{\frac{1}{2}}\hat{\varkappa}_3-\Omega\left(\frac{A}{\Delta\Sigma}\right)^{\frac{1}{2}}\hat{\varkappa}_0
\label{inv}
\end{eqnarray}
we write (\ref{deta}) as,
 \begin{eqnarray}
d\eta=\lambda_{10}\hat{\varkappa}_1\wedge\hat{\varkappa}_0+\lambda_{20}\hat{\varkappa}_2\wedge\hat{\varkappa}_0+\lambda_{13}\hat{\varkappa}_1\wedge\hat{\varkappa}_3+\lambda_{23}\hat{\varkappa}_2\wedge\hat{\varkappa}_3
\label{eta11}
\end{eqnarray}
where
\begin{eqnarray}
\lambda_{10}&=&-\frac{\partial g_{03}}{\partial r}\frac{A^{\frac{1}{2}}}{\Sigma}-\frac{\partial g_{33}}{\partial r}\Omega\frac{A^{\frac{1}{2}}}{\Sigma}\nonumber\\
\lambda_{20}&=&\frac{\partial g_{03}}{\partial \theta}\frac{1}{\Sigma}\left(\frac{A}{\Delta}\right)^{\frac{1}{2}}-\frac{\partial g_{33}}{\partial \theta}\frac{\Omega}{\Sigma}\left(\frac{A}{\Delta}\right)^{\frac{1}{2}}\nonumber\\
\lambda_{13}&=&\frac{\partial g_{33}}{\partial r}\frac{1}{{\textrm{sin}}\theta}\left(\frac{\Delta}{A}\right)^{\frac{1}{2}}\nonumber\\
\lambda_{23}&=&\frac{\partial g_{33}}{\partial \theta}\frac{1}{{\textrm{sin}}\theta}\left(\frac{1}{A}\right)^{\frac{1}{2}}
\label{lam}
\end{eqnarray}
 The dual of (\ref{eta11}) is 
\begin{eqnarray}
^* d\eta=-\lambda_{10}\hat{\varkappa}_2\wedge\hat{\varkappa}_3+\lambda_{20}\hat{\varkappa}_1\wedge\hat{\varkappa}_3+\lambda_{13}\hat{\varkappa}_2\wedge\hat{\varkappa}_0-\lambda_{23}\hat{\varkappa}_1\wedge\hat{\varkappa}_0
\label{}
\end{eqnarray}
Using (\ref{ford}), above equation is written in usual coordinate two form as,
\begin{eqnarray}
^* d\eta =\delta_{rt}dr\wedge dt+\delta_{\theta t}d\theta\wedge dt+\delta_{r\phi}dr\wedge d\phi+\delta_{\theta\phi}d\theta\wedge d\phi
\label{eta1}
\end{eqnarray}
where
\begin{eqnarray}
\delta_{rt}&=&-\lambda_{20}\Omega\left(\frac{A{\textrm{sin}}^2\theta}{\Delta}\right)^{\frac{1}{2}}+\lambda_{23}\Sigma\left(\frac{1}{A}\right)^{\frac{1}{2}} \nonumber\\
\delta_{\theta t}&=&-[\lambda_{13}\Sigma\left(\frac{\Delta}{A}\right)^{\frac{1}{2}}- \lambda_{10}\Omega\sqrt{A}{\textrm{sin}}\theta]\nonumber\\
\delta_{r\phi}&=&\lambda_{20}\left(\frac{A{\textrm{sin}}^2\theta}{\Delta}\right)^{\frac{1}{2}}\nonumber\\
\delta_{\theta\phi}&=&-\lambda_{10}\sqrt{A}{\textrm{sin}}\theta
\label{delta}
\end{eqnarray}

To calculate the effective Komar angular momentum (\ref{peff}) one needs to choose an appropriate boundary surface ($\partial\Sigma$). It is the boundary of a spatial three volume ($\Sigma$) characterised by a constant $r$ and $dt=-\frac{g_{03}}{g_{00}}d\phi$. Under this condition (\ref{eta1}) is simplified as
\begin{eqnarray}
^*d\eta =-\frac{g_{03}}{g_{00}}\delta_{\theta t}d\theta\wedge d\phi+\delta_{\theta\phi}d\theta\wedge d\phi
\label{aaaa}
\end{eqnarray}
Now substituting (\ref{aaaa}) in (\ref{peff}), we find the expression of Komar angular momentum as,\begin{eqnarray}
K_{\xi^{\mu}_{(\phi)}}=\frac{1}{8\pi}\int{\frac{g_{03}}{g_{00}}\delta_{\theta t}d\theta d\phi}-\frac{1}{8\pi}\int{\delta_{\theta\phi}d\theta d\phi}
\label{komas1}
\end{eqnarray}
Moving along a closed contour, the first term of the right hand side gives the shift of time between the initial and the final events. Since we are performing an integration over simultaneous events this term must be subtracted from (\ref{komas1}) \cite{cohen1,cohen2}.  So we write the above equation as,
\begin{eqnarray}
K_{\xi^{\mu}_{(\phi)}}=\frac{1}{8\pi}\int \lambda_{10}\sqrt{A}{\textrm{sin}}\theta d\theta d\phi
\label{komas}
\end{eqnarray}
where we have used the relation (\ref{delta}). Substituting $\lambda_{10}$ from (\ref{lam}), we write the above expression as
\begin{eqnarray}
K_{\xi^{\mu}_{(\phi)}}=\frac{1}{8\pi}\int\left(-\Omega\frac{\partial g_{33}}{\partial r}-\frac{\partial g_{03}}{\partial r}\right)\frac{A{\textrm{sin}}\theta}{\Sigma}d\theta d\phi
\label{}
\end{eqnarray}
Using the metric coefficients
\begin{eqnarray}
g_{03}&=&\frac{a(Q^2-2Mr){\textrm{sin}}^2\theta}{r^2+a^2{\textrm{cos}}^2\theta}\\
g_{33}&=&\frac{{\textrm{sin}}^2\theta\left((a^2+r^2)(r^2+a^2{\textrm{cos}}^2\theta)-a^2(Q^2-2Mr){\textrm{sin}}^2\theta\right)}{r^2+a^2{\textrm{cos}}^2\theta}
\label{}
\end{eqnarray}
we find,
\begin{eqnarray}
K_{\xi^{\mu}_{(\phi)}}&=&\frac{1}{4}\int\frac{2a}{(r^2+a^2{\textrm{cos}}^2\theta)^2}[r^3(2Q^2-3Mr)+ra^2(Q^2-Mr)+\nonumber\\&&a^2(a^2M+r(Q^2-Mr)){\textrm{cos}}^2\theta]{\textrm{sin}}^3\theta d\theta
\label{}
\end{eqnarray} 
After performing the integration we find the Komar conserved quantity corresponding to the spacelike one form $\eta$, given by, 
\begin{eqnarray}
K_{\xi^{\mu}_{(\phi)}}=-\left(2aM+\frac{rQ^2}{2a}-\frac{aQ^2}{2r}-\frac{Q^2(a^2+r^2)^2}{2a^2r^2}{\textrm{tan}}^{-1}(\frac{a}{r})\right)
\label{jeff}
\end{eqnarray}
Taking the asymptotic limit one finds $\displaystyle\lim_{r\rightarrow\infty}~K_{\xi^{\mu}_{(\phi)}}=-2Ma$. This is different from the angular momentum of Kerr-Newman black hole $J=Ma$ {\footnote{For a possible justification for the anomalous factor $2$ see \cite{katz}.}}. Therfore in order to derive the correct expression of angular momentum the normalisation of (\ref{peff}) should be changed accordingly. Thus our result (\ref{jeff}) is divided by $-2$ to get the correct result of effective angular moentum, which yields,
\begin{eqnarray}
J_{\textrm{eff}}=Ma+\frac{rQ^2}{4a}-\frac{aQ^2}{4r}-\frac{Q^2(a^2+r^2)^2}{4a^2r^2}{\textrm{tan}}^{-1}(\frac{a}{r})
\label{jeffc}
\end{eqnarray}
This is a new result and shows at finite radial distance the angular momentum of the Kerr-Newman black hole gets modified from its asymptotic value $J=Ma$. In fact similar to $M_{\textrm{eff}}$ in (\ref{meff}), $J_{\textrm{eff}}$ also is always less than its asymptotic value. All the extra contributions appearing here are due to the presence of electromagnetic matter field outside the black hole event horizon. Note that in the chargeless case all other terms vanish and therefore one can conclude that for rotating Kerr spacetime angular momentum is unchanged with the variation of radial distance. In the irrotational limit, {\it i.e.}  $a=0$, the expression (\ref{jeffc}) vanishes which is reassuring.     

  
\section{Application of effective Komar conserved quantities}
After getting the expressions of Komar conserved quantities (\ref{meff},\ref{jeff}) for the Kerr-Newman spacetime we now use them to derive some important results at the black hole event horizon ($r=r_+$). This will enable us to understand some fetures of black hole physics in a transparent manner.
\subsection{The identity $K_{\chi^{\mu}}=2ST$}
We have already mentioned that the Kerr-Newman spacetime is an axially symmetric spacetime and it has two Killing vectors. These two vectors $\xi^{\mu}_{(t)}=(1,0,0,0)$ and $\xi^{\mu}_{(\phi)}=(0,0,0,1)$ are respectively timelike and spacelike in the asymptotic limit. Although none of these two vectors are globally hypersurface orthogonal timelike vectors, the combination of these two vectors $\chi^{\mu}=\xi^{\mu}_{(t)}+\Omega\xi^{\mu}_{(\phi)}=(1,0,0,\Omega)$ is timelike in the entire spacetime (outside the event horizon) \cite{Carrol, poisson}. However this is not a Killing vector for arbitrary $r$ because of the fact that the angular velocity ($\Omega$), even for a fixed $r$ depends upon the polar angle $\theta$. Only at the event horizon, $r=r_+=M+\sqrt{M^2-a^2-Q^2}$, $\Omega=\Omega_{\textrm{H}}=\frac{a}{r_+^2+a^2}$ is a constant and $\chi^{\mu}\chi_{\mu}|_{r_+}=0$, thus $\chi^{\mu}$ becomes a null Killing vector. Thus the event horizon of the Kerr-Newman black hole is also a Killing horizon with respect to the Killing vector, 
\begin{eqnarray}
\chi^{\mu}=\xi^{\mu}_{(t)}+\Omega_H\xi^{\mu}_{(\phi)}
\label{neew}
\end{eqnarray}

Now we shall use (\ref{meff}) and (\ref{jeff}) to derive a new identity at the black hole event horizon. Since $\chi^{\mu}$ is Killing at the event horizon we can define the Komar conserved quantity corresponding to this vector, given by 
\begin{eqnarray}
K_{\chi^{\mu}}=-\frac{1}{8\pi}\int_{\partial \Sigma}{}^*d\chi.
\end{eqnarray}
Using (\ref{neew}) this can be written as
\begin{eqnarray}
K_{\chi^{\mu}}=K_{\xi^{\mu}_{(t)}}+\Omega_H K_{\xi^{\mu}_{(\phi)}}.
\label{efen1}
\end{eqnarray}
Using the expression $\Omega_{\textrm{H}}=\frac{a}{r_+^2+a^2}$ together with (\ref{meff}) and (\ref{jeff}) in the above equation we find,
\begin{eqnarray}
K_{\chi_{\mu}}&=&M-2\frac{a^2}{r_+^2+a^2}M-\frac{Q^2r_+}{r_+^2+a^2}
\label{efen2}
\end{eqnarray}
Using the expression of $r_+(r_+=M+\sqrt{M^2-a^2-Q^2})$ we simplify the above equation to write it in the form
\begin{eqnarray}
K_{\chi_{\mu}}&=&\sqrt{M^2-a^2-Q^2}\\
&=&2\left(\pi\left(2Mr_+-Q^2\right)\right)\left(\frac{\sqrt{M^2-a^2-Q^2}}{2\pi\left(2Mr_+-Q^2\right)}\right)
\label{efen222}
\end{eqnarray}
Since the expressions for entropy and temperature for this black hole are
\begin{eqnarray}
S&=&\pi\left(2Mr_+-Q^2\right)\\
T&=&\frac{\sqrt{M^2-a^2-Q^2}}{2\pi\left(2Mr_+-Q^2\right)}
\end{eqnarray}
we finally obtain the identity
\begin{eqnarray}
K_{\chi_{\mu}}&=&2ST
\label{iden11}
\end{eqnarray}
This result expresses the fact that at the event horizon the Komar conserved charge corresponding to the null Killing vector is twice the product of temperature and entropy of the Kerr-Newman black hole. Such a relation for this black hole was also derived  in \cite{Banerjee:2010yd}  by using a different method. 

To connect our result with the recent studies on the thermodynamic interpretation of gravity we cite the works \cite{paddy1,paddy2,Padmanabhan:2009vy,Banerjee:2010yd,Banerjee:2010rx} where it has been shown that an  identity $E=2ST$ is always satisfied at the Killing horizon of a spacetime. Here $E$ is the conserved Noether charge corresponding to a diffeomorphic transformation of the metric. This relation has been proved particularly for static local Killing horizon \cite{Padmanabhan:2009vy} which may or may not be a black hole event horizon. However Kerr-Newman spacetime is nonstatic and therefore it is not clear whether such a relation is still valid there or not. Nevertheless here we have explicitly derived a similar relationship at the event horizon of Kerr-Newman black hole and showed that the left hand side of the identity is directly associated with the Komar conserved charges. 

To understand the implication of the above identity we recall (\ref{efen2}). Since the scalar potential $\Phi$ at the event horizon is given by,
\begin{eqnarray}
\Phi=\frac{Qr_+}{r_+^2+a^2}
\end{eqnarray}
(\ref{efen2}) is written as
\begin{eqnarray}
K_{\chi_{\mu}}=M-2\Omega_H J-Q\Phi
\label{efen3}
\end{eqnarray}
In general, the temperature and entropy of a black hole can be written in terms of its surface gravity ($\kappa$) and horizon area ($A_{\text{Horizon}}$) as $T=\frac{\kappa}{2\pi}$ and $S=\frac{A}{4}$. Using these results together with (\ref{efen3}) and the identity (\ref{iden11}) we get
\begin{eqnarray}
M-2\Omega_H J-Q\Phi=\frac{\kappa A_{\text{Horizon}}}{4\pi}
\label{smf}
\end{eqnarray}
This well known result is usually referred as the Smarr mass formula. In the literature there are two well known methods of deriving this formula. One of which is due to the original work of Smarr \cite{smarr} and the other one is proposed by Bardeen, Carter and Hawking \cite{Bardeen}. Among these the first one \cite{smarr} is a metric dependent approach based on the concepts of Euler's theorem of homogenous function and the other one is metric independent approach \cite{Bardeen}. In our method this formula is a direct consequence of the identiy (\ref{iden11}) whose root lies in the Komar integral invariants associated with various Killing symmetries. 

Note that, the relations (\ref{iden11}) and (\ref{smf}) are not only different manifestations of an equation, but also they are completely different with respect to its interpretation. While (\ref{iden11}) is an identity defined exactly at the event horizon, (\ref{smf}) includes some physical parameters which are defined at distinct regions of the spacetime. For example, in (\ref{smf}) angular velocity ($\Omega_{\text{H}}$), surface gravity ($\kappa$), horizon area ($A_{\text{Horizon}}$) and electric potential ($\Phi$) are defined at the black hole event horizon. On the other hand mass ($M$) and the angular momentum ($J$) are defined at the asymptotic infinity. Thus Smarr formula, as written in (\ref{smf}), is a non-local relationship. And the identity (\ref{iden11}) can be interpreted as a local version (\ref{smf}).      

\subsection{Leading correction to black hole entropy}
It is interesting to see the use of effective Komar conserved quantities to find the higher order corrections to the semiclassical Bekenstein-Hawking entropy. The corrected entropy of the Kerr-Newman black hole, as found in a recent work \cite{Banerjee:2009tz} involving one of us, is given by
\begin{eqnarray}
S_{\text{cor}}=S+2\pi\beta_1\log{S}-\frac{4\pi^2\beta_2}{S}+{\cal O}(\hbar^2).
\label{coren}
\end{eqnarray}
By making an infinitesimal scale transformation to the black hole metric the unknown coefiicients ($\beta_i$) are found to be related with the trace anomaly (${\langle T^{\mu}_{\mu}\rangle}^{i}$) of the energy momentum tensor . Particularly the coefficient of the leading (logarithmic) correction is given by
\begin{eqnarray}
\beta_1=-\frac{\sqrt{M^2-a^2-Q^2}}{4\pi\omega}{\textrm{Im}}\int{\sqrt{-g}{\langle{T^\mu~_{\mu}}\rangle}^{(1)}}.
\label{coeff}
\end{eqnarray}
Here $\omega=K_{\xi^{\mu}_{(t)}}|_{r=r_+}+\Omega_H K_{\xi^{\mu}_{(\phi)}}|_{r=r_+}$ is the effective energy at the event horizon and ${\langle{T^\mu~_{\mu}}\rangle}^{(1)}=\frac{1}{2880\pi^2}(R_{\mu\nu\rho\sigma}R^{\mu\nu\rho\sigma}-R_{\mu\nu}R^{\mu\nu}+\nabla_{\mu}\nabla^{\mu}R)$ is the trace anomaly of the scalar field in one loop calculation. To calculate $\beta_1$ we need to find $\omega$. Using (\ref{meff},\ref{jeff}) we find $\omega=\sqrt{M^2-a^2-Q^2}$. In \cite{Banerjee:2009tz} we found the same result for $\omega$ by using an approximation while in the present case it is evaluated exactly. Finally substituting this value of $\omega$ in (\ref{coeff}) and integrating over the trace anomaly one can find $\beta_1$ \cite{Banerjee:2009tz} and the leading correction to the black hole entropy.

\section{Conclusions}
In this paper we found the effective angular momentum at any distance $r$ for Kerr-Newman (KN) spacetime. This was obtained by evaluating the Komar conserved expressions at the boundary of a finite spatial surface of radius $r$. This boundary was chosen in such a manner that it only contains the time synchronized events. This choice enabled us to evaluate the Komar integrals without any asymptotic approximation. Using these effective expressions we studied some important features at the black hole event horizon (EH). We found the null Killing vector ($\chi^{\mu}$) at the EH and calculated the Komar conserved charge ($K_{\chi^{\mu}}$) corresponding to this vector. Interestingly this conserved charge was found to be twice the product of entropy and temperature of the black hole ($K_{\chi^{\mu}}=2ST$). Since KN space time is nonstatic in nature this identity is a generalization of a similar relation $E=2ST$ which is known to be true for a static spacetime. Thus in the present paper we made some progress in understanding the relation between geometric and thermodynamic quantities of a spacetime. As an application of this identity we derived the generalized Smarr formula for KN black hole in a very natural manner. We also argued that the identity derived in this paper can be interpreted as a local form of the non-local Smarr mass formula. Finally we used the Komar effective expressions to evaluate the coefficient of the leading correction to black hole entropy. 

\section*{Acknowledgement} 
One of the authors (S.K.M) thanks the Council of Scientific and Industrial Research (C.S.I.R), Government of India, for financial support.


\end{document}